**Targeting glutamate metabolism in melanoma**

*Simar Singh [1,2,3 $], Raj Shah [4,5 $], Suzie Chen [4,5,6], Fabian V. Filipp [1,2]*

[1] Cancer Systems Biology, Institute of Computational Biology, Helmholtz Zentrum München, Ingolstädter Landstraße 1, D-85764, München, Germany.

[2] School of Life Sciences Weihenstephan, Technical University München, Maximus-von-Imhof-Forum 3, D-85354, Freising, Germany.

[3] St. George's University, School of Medicine, Grenada, West Indies.

[4] Joint Graduate Program in Toxicology, Rutgers University, Piscataway, New Jersey.

[5] Susan Lehman Cullman Laboratory for Cancer Research, Ernest Mario School of Pharmacy, Department of Chemical Biology, Rutgers University, Piscataway, New Jersey.

[6] Rutgers Cancer Institute of New Jersey, New Brunswick, New Jersey.

[$] **shared contribution**

Fabian V. Filipp fabian.filipp@helmholtz-muenchen.de ; Suzie Chen suziec@pharmacy.rutgers.edu

**Abstract**

The glutamate metabotropic receptor 1 (GRM1) drives oncogenesis when aberrantly activated in melanoma and several other cancers.

Metabolomics reveals that patient-derived xenografts with GRM1-positive melanoma tumors exhibit elevated plasma glutamate levels associated with metastatic melanoma *in vivo*. Stable isotope tracing and GCMS analysis determined that cells expressing GRM1 fuel a substantial fraction of glutamate from glycolytic carbon. Stimulation of GRM1 by glutamate leads to activation of mitogenic signaling pathways, which in turn increases the production of glutamate, fueling autocrine feedback. Implementing a rational drug-targeting strategy, we critically evaluate metabolic bottlenecks *in vitro* and *in vivo*.

Combined inhibition of glutamate secretion and biosynthesis is an effective rational drug targeting strategy suppressing tumor growth and restricting tumor bioavailability of glutamate.

**Introduction**

The metabotropic glutamate receptors (GRMs) are seven pass transmembrane G-protein-coupled-receptors (GPCR) that elicit intracellular signaling upon binding extracellular glutamate [1-4]. Though well characterized in the context of excitatory neurotransmission in the central nervous system, GRM-mediated signaling is now understood to play a role in several neuronal and non-neuronal pathologies, including cancer [5-9]. Elevated tumor expression of GRMs has been reported not only in central nervous system tumors where glutamate is abundant, but also in melanoma, breast, and prostate cancers [10-13]. The neuroectodermal origin of melanocytes, pigment producing cells that migrated and differentiated from the neural crest into the skin, might be of particular relevance in malignant melanoma [14]. Metastases to the central nervous system are among the most common and lethal complications of metastatic melanoma. Melanoma is the third most common source of brain metastases following lung and breast cancer and more than 60% of patients with metastatic melanoma either present with or develop brain metastases during the course of their disease.

The first direct link between the glutamate metabotropic receptor 1 (GRM1) and oncogenesis came from a transgenic mouse model of melanoma ectopically expressing GRM1 [15-17]. Since then, it has been reported that 80% of human melanoma cell lines and 65% of primary and secondary melanoma biopsies express GRM1 [10]. GRM1 overexpression is sufficient to induce oncogenic transformation, and pharmacological inhibition of GRM1 in high expressing tumors slows proliferation and tumor progression [10,15,18-21]. Stimulation of GRM1 signaling by glutamate or other agonists results in activation of oncogenic signaling pathways including mitogen activated protein kinase (MAPK), phosphatidylinositol 3-kinase (PI3K) and serine/threonine kinase AKT, and insulin-like growth factor 1 (IGF1) [10,15,22-24]. These and other results indicate that GRM1 is an oncogene that may serve as an attractive therapeutic target [25], especially in melanoma subtypes, where well-characterized drug targets beyond the MAPK pathway are missing. Metabolic rewiring, including increased glucose and glutamine consumption, is a prominent feature of malignant melanoma [26-29]. GRM1 expressing cells increase the production and release of glutamate, thereby engaging in an autocrine loop involving leading to increased GRM1 mediated oncogenic signaling [10,30-32]. Given metabolic similarity of the five-carbon amino acids glutamate and glutamine and their relatively direct conversion by deamination, the L-glutamine amidohydrolase, glutaminase (GLS), has been suggested as carbon bottleneck and anti-cancer drug target.

Treatment with riluzole, which blocks glutamate release, disrupts GRM1 mediated signaling and exhibits anti-tumor properties. Pharmacologically, bioavailability and molecular targeting of glutamate metabolism are active areas of research, despite the exact mechanism of how riluzole inhibits the release of glutamate is biochemically unknown. Preclinical studies with riluzole, which serves as an indirect antagonist of GRM1 and blocks glutamate release, led to decreased melanoma cell growth *in vitro* and tumor progression *in vivo* [19 33-37]. Understanding the metabolic pathways employed by GRM1-activated melanoma to increase glutamate production may lead to the development of rational drug combinations targeting aberrant GRM1 signaling and its accompanying metabolic rewiring. With applicable experimental cellular and animal model systems in hand (Figure 1), we sought to elucidate the metabolic source of activated glutamate in GRM-activated tumors and to design a rational drug targeting strategy of glutamate metabolism in malignant melanoma.

## Methods

*Constructing isogenic melanoma models with modulated GRM1 expression*

We metabolically profiled circulating blood from tumor-bearing xenograft animals as well as the source panel of isogenic human melanoma cell lines engineered to have increased or suppressed GRM1. Patient-derived human melanoma cell lines were provided by Dr. Mary J.C. Hendrix (Children's Memorial Research Center, Chicago, IL). C81-61 is an early stage melanoma cell line, which does not express endogenous GRM1. C8161 is a malignant metastatic melanoma cell line from the same patient as C81-61 but with hyperactive GRM1. A stable C8161 TetR GRM1 siRNA knockdown B22-20 clone was generated and maintained in 1 µg/ml blasticidin and 10 µg/ml hygromycin [19]. Induction of siRNA against GRM1 was carried out by incubating the cells with 10 ng/ml of doxycycline for 4 days. Stable C81-61OE GRM1-6 clone that expresses elevated GRM1 levels compared to parental cell lines was selected with 10 µg/ml blasticidin [38]. These cell lines were cultured in DMEM (10-017, Corning Cell-Gro, Manassas, VA) supplemented with 10% fetal bovine serum (FBS), 1% Penicillin-Streptomycin (30-002-CI, Corning Cell-Gro, Manassas, VA), and 1% MEM Non-Essential Amino Acids (25-025-C, Corning Cell-Gro, Manassas, VA) at 37.0 °C (310.15 K) with 5.0% carbon dioxide (CO2, CD50, Praxair, Danbury, CT). Together, the panel of human cell lines, C81-61 ($GRM1^{-}$), C8161si TetR siRNA(*GRM1*) ($GRM1^{KD}$), C81-61OE ($GRM1^{OE}$), and C8161 ($GRM1^{+}$) provide a patient-derived isogenic progression model of NF1/NRAS/BRAF triple wild-type melanoma (Figure 1).

The 1205LU (ATCC CRL-2812) model was provided by Dr. Meenhard Herlyn (Wistar Institute, Philadelphia, PA) and is derived from lung metastases of primary melanoma lesion WM793B cells (ATCC CRL-2806) after subcutaneous injection into immunodeficient mice. 1205LU cells are highly invasive and exhibit spontaneous metastasis to lung and liver [39]. The WM793B line was established from skin taken from the primary melanoma of a vertical growth phase (VGP) lesion taken from the sternum of a patient on 01/07/1983.

The transgenic founder strain 3 (TG3) with hyperactive *Grm1*$^{+}$ was crossed with hairless SKH-1 mice to arrive at the TGS strain—brother-sister littermates have been mating since the year 2000. TG3 mice were established as a result of a classic case of insertional mutagenesis that led to the ectopic expression of Grm1 in melanocytes. TG3 mice spontaneously develop metastatic melanoma with 100% penetrance. SKH-1 is an uncharacterized/non-pedigreed hairless strain of mice. The goal was to make the pigmented lesions visible on the TGS mice in the absence of fur. TGS mice were genotyped by DNA extraction and subsequent PCR of the *Grm1* locus.

*Toxicology and inhibitor studies*

The inhibitory concentration 50% (IC50), the concentration of drug that is required for 50% inhibition of cell proliferation *in vitro*, was determined in 384 plate-based cell proliferation assays at 450 nm absorption. After 24 h incubation, cells were equilibrated at room temperature before being treated with 50 µL/well of CellTiter-Glo luminescent cell viability assay (CTG, G7571, Promega, Madison, WI). The number of viable cells relative to control cells, results in an accompanying change in the amount of formazan formed using a saturated aqueous 1.0 mg/mL solution of 2,3-bis[2-methoxy-4-nitro-5-sulfophenyl]2H-tetrazolium-5-carboxyanilide sodium salt (XTT, X4626, Millipore Sigma, Darmstadt, Germany), indicating electron-dependent redox changes or the degree of cytotoxicity caused by the test compounds. The compounds were dissolved as stock solutions in dimethyl sulfoxide (DMSO, D2650, Millipore Sigma, Darmstadt, Germany) or water and compared to the vehicle-treated control conditions. Cells were plated in black walled, clear bottom 384 well plates at a density of 500 cells per well in 50 µL of supplemented MEM. After 24 h, 5 µL of a 2-fold dilution series of each chemical inhibitor in supplemented MEM was added to wells in quadruplicate (N=4). Cells were then shaken for 2.0 min and incubated for 8.0 min protected from light before the luminescence is being analyzed in a multi-mode microplate reader (Synergy HT, BioTek, Winooski, VT).

Tested compounds included Bis-2-(5-phenylacetamido-1,3,4-thiadiazol-2-yl)ethyl sulfide; 2,2'-(5,5'-(2,2'-thiobis(ethane-2,1-diyl))bis(1,3,4-thiadiazole-5,2-diyl))bis(azanediyl)bis(1-phenylethanone (BPTES, PubChem CID: 3372016, SML0601), deferoxamine mesylate (PubChem CID: 62881, D9533), sodium oxamate (PubChem CID: 5242, O2751), dichloroacetic acid (DCA, PubChem CID: 6597, 347795), 5-aminoimidazole-4-carboxamide 1-beta-D-ribofuranoside (AICAR, PubChem CID: 46780289, A9978), 2-deoxy-D-glucose (2DG, D6134) and 2-cyano-3-(1-phenylindol-3-yl)acrylate (UK5099, PubChem CID: 6438504, PZ0160, all Millipore Sigma, Darmstadt, Germany), 2-(pyridin-2-yl)-n-(5-(4-[6-(([3-(trifluoromethoxy)phenyl]acetyl)amino)pyridazin-3-yl]butyl)-1,3,4-thiadiazol-2-yl)acetamide (CB839, PubChem CID: 71577426, Medchemexpress, Monmouth Junction, NJ), and 6-(trifluoromethoxy)-1,3-benzothiazol-2-amine (riluzole, PubChem CID: 5070, Selleckchem, Houston, TX).

*Xenograft studies and tumorigenicity assays*

C8161 and 1205LU human melanoma cells were harvested by trypsinization and resuspended in phosphate based saline (PBS) ($10^7$ cells/ml). 5 to 6-week-old immunodeficient nude male and female mice were subcutaneously injected with $10^6$ tumor cells in each dorsal flank. Tumor cell lines for xenograft studies were cultured in RPMI-1640 medium supplemented with 10% FBS. Tumor growth was monitored weekly with a vernier caliper and calculated with the formula ($d^2 \cdot D/2$). CB839 (50 mM stock solution), riluzole (100 mM stock solution) and UK5099 (50 mM stock solution) were dissolved in DMSO. Once tumor volumes reached 10 to 20 $mm^3$, mice were randomly divided into four treatment groups and received control vehicle DMSO, riluzole (10 mg/kg), CB839 (200 mg/kg), or the combination of riluzole (10 mg/kg) and CB839 (200 mg/kg) by oral gavage, daily. UK5099 was administered via intraperitoneal injection (3 mg/kg). The experiment was terminated when the xenografts in the vehicle group reached maximum permitted size. All animal procedures and xenograft studies were performed in strict accordance with the institutional animal care and use committee (IACUC).

*Blood collection*

Heparinized microhematocrit capillary tubes (22-362-566, Thermo Fisher Scientific, Waltham, MA) were used to retro-orbitally bleed mice. About 300 µL of blood was collected per mouse at each time-point and kept on ice. Samples were centrifuged for 8 min at $10^4$ rpm at 4 °C (277.15 K). Plasma supernatant was separated from the red blood cell pellet and stored at -80 °C (193.15 K).

*Immunohistochemistry*

Sectioning, staining and quantitative image analysis of formalin-fixed, paraffin-embedded (FFPE) tumor and liver tissues were performed using a fully automated workflow according to standard operating procedures (Histowiz, Brooklyn, NY). Briefly, tissue samples were processed and embedded in paraffin blocks and then shipped out to the pathology lab on ice where subsequent sectioning at 4 µm and Immunohistochemistry (IHC) was piloted. FFPE liver sections were stained with hematoxylin and eosin (H&E). IHC on FFPE tumor tissue was performed on a Bond Rx autostainer (Leica Biosystems, Nussloch, Germany) with enzyme treatment (1:1000) using standard protocols. Bond polymer refine detection (Leica Biosystems, Nussloch, Germany) was used according to standard operating procedures. After staining, sections were dehydrated and film cover-slipped using a TissueTek-Prisma and Coverslipper (Sakura Finetek, Torrance, CA). Whole slide scanning (40x magnification) was performed on an Aperio AT2 (Leica Biosystems, Nussloch, Germany). Unbiased automated image analysis to obtain percentage of marker of proliferation Ki-$67^+$ (MKI67) and cleaved apoptosis marker executioner caspase $3^+$ (CASP3) cells was performed using the HALO image analysis software by Indica Labs, Albuquerque, NM).

*Western immunoblots*

Polyclonal anti-GRM1 antibody was purchased from Lifespan BioSciences (LS-C354444, Seattle, WA). Polyclonal anti-GLS antibody was purchased from Novus Biologicals (NBP1-58044, Littelton, CO). Monoclonal anti-α-tubulin antibody was purchased from Sigma Aldrich (T6074, St. Louis, MO). Monoclonal anti-phospho-T308-AKT (4056), pan-AKT1/2/3 (4691), phospho-ERK1/2-T202/Y204 (4370), pan-ERK1/2 (4695), and anti-PARP (9532) antibodies were purchased from Cell Signaling Technology (Danvers, MA). All primary antibodies recognize human proteins and were produced in rabbit. A peroxidase conjugate served as secondary goat anti-mouse IgG antibody (A4416, Millipore Sigma, Darmstadt, Germany). Cell lines were maintained in a humidified 5.0% $CO_2$ incubators at 37.0 °C (310.15 K). In order to prepare whole cell protein lysates, media was removed and cells were washed once with ice-cold PBS. After removal of PBS, extraction buffer was added directly to the plates and cells were collected with a cell scraper. Cell extracts were incubated on ice for 20 min. Cell debris was removed by centrifugation at 14,000 rpm at 4 °C (277.15 K) for 20 min, supernatant was collected and stored at -80 °C (193.15 K). 25 µg of protein was routinely used for Western immunoblot analysis. The intensity of the protein bands on the blots was quantified using National Institutes of Health-approved ImageJ platform (National Institutes of Health, Bethesda, MA).

*Statistical analysis*

Statistical significance of experimental data was evaluated by a Student's t-test or Fisher analysis of variance. The statistical significance thresholds was set at *p* values of *p<0.05, **p<0.01 or ***p<0.001.

*L-glutamine deprivation and glutaminase inhibition*

Cells were plated in white 96-well assay plates at a density of $10^4$ cells/well in 100 µL of media containing either no L-glutamine, 2.0 mM L-glutamine, or 2.0 mM L-glutamine with 1.0 mM selective glutaminase inhibitor bis-2-(5-phenylacetamido-1,3,4-thiadiazol-2-yl)ethyl sulfide (BPTES, SML0601, Sigma Merck, Darmstadt, Germany). After 24 h, cells were equilibrated at room temperature before being treated with 100 µL/well of CellTiter-Glo luminescent cell viability assay and analyzed.

*Cell culture for metabolomics measurements*

For metabolite quantification and stable isotope tracing, $10^6$ cells per well were seeded in replicate (N=6) in 6 well plates (657160, Greiner Bio-One, Kremsmünster, Germany) in DMEM (10-017, Corning Cell-Gro, Manassas, VA) supplemented with 10% FBS, 1% Penicillin-Streptomycin (30-002-CI, Corning Cell-Gro, Manassas, VA), and 1% MEM Non-Essential Amino Acids (25-025-C, Corning Cell-Gro, Manassas, VA). Following a 24 h seeding, attachment, and equilibration period, media was aspirated and replaced with MEM (Corning Cell-Gro, Manassas, VA) supplemented with 1.0 g/L D-glucose (0188, Amresco, Solon, OH), 2.0 mM L-Glutamine (G3126, Sigma Merck, Darmstadt, Germany), 10% FBS, and 1% MEM Vitamins (25-020-CI, Corning Cell-Gro, Manassas, VA). For glucose and glutamine labeling experiments, MEM media was supplemented with 1.0 g/L [U-$^{13}C_6$] D-glucose ([U-13C6] GLC, 389374-2G, Sigma Isotec, Miamisburg, OH) or 2.0 mM [U-13C5] L-glutamine ([U-$^{13}C_5$] GLN, 605166-500MG, Sigma Isotec, Miamisburg, OH).

*Metabolite extraction*

Following 24 h incubation in supplemented MEM, 5 µL of supernatant containing blood plasma, cellular samples, or conditioned media was transferred to microcentrifuge tubes (MT-0200-BC, Biotix, San Diego, CA) with 1 mL of cold -20 °C (253.15 K) extraction buffer consisting of 50% methanol (A452, Fisher Scientific, Fair Lawn, NJ) in ultrapure (18.2 MΩ x cm) water with 20µM L-norvaline (N7627 Sigma Merck, Darmstadt, Germany) and 20µM DL-norleucine (N1398, Sigma Merck, Darmstadt, Germany) and dried by vacuum centrifugation in a speedvac concentrator (DNA120OP115, Savant, Thermo Fisher Scientific, Waltham, MA) overnight. The remaining media was aspirated and the cells washed quickly with cold 0.9% sodium chloride in ultrapure water (Amresco, Solon, OH) and placed on ice. To each well, 1 mL of cold extraction buffer was added, the cells were collected on ice using a cell scraper and the entire solution was then transferred to a pre-chilled microcentrifuge tube. Tubes were then frozen in liquid nitrogen, thawed, and placed in a digital shaking drybath (8888-0027, Thermo Fisher Scientific, Waltham, MA) set to 1100rpm for 15 min at 4 °C (277.15 K). Samples were then centrifuged for 15 min at 4 °C (277.15 K) and 12500 g in a refrigerated centrifuge (X1R Legend, Sorvall, Thermo Fisher Scientific, Waltham, MA) using a fixed-angle rotor (F21-48x1.5, Sorvall, Thermo Fisher Scientific, Waltham, MA). Supernatants were transferred to new microcentrifuge tubes and dried by vacuum centrifugation overnight.

*Metabolite derivatization*

Dried, extracted plasma, cell samples or media supernatants were derivatized by addition of 20 µL of 2.0% methoxyamine-hydrochloride in pyridine (MOX, TS-45950, Thermo Fisher Scientific, Waltham, MA) followed by a 90 min incubation period in a digital shaking drybath at 30 °C (303.15 K) and 1100 rpm. 90 µL of N-methyl-N-(trimethylsilyltrifluoroacetamide (MSTFA, 394866-10X1ML, Sigma Merck, Darmstadt, Germany) was added and samples incubated at 37 °C (310.15 K) and 1100 rpm for 30 min before centrifugation for 5 min at 14,000 rpm and 4 °C (277.15 K). The supernatant was transferred to an autosampler vial (C4000LV3W, Thermo Fisher Scientific, Waltham, MA) with screwcap (C5000-53B, Thermo Fisher Scientific, Waltham, MA) for separation by gas chromatography (GC, TRACE 1310, Thermo Fisher Scientific, San Jose, CA) coupled to a triple-quadrupole GC mass spectrometry system for analysis (QQQ GCMS, TSQ8000EI, TSQ8140403, Thermo Fisher Scientific, San Jose, CA).

*Metabolomics separation and detection*

Samples were analyzed on a QQQ GCMS system equipped with a 0.25 mm inner diameter, 0.25 µm film thickness, 30 m length 5% diphenyl/95% dimethyl polysiloxane capillary column (OPTIMA 5 MS Accent, 725820.30, Machery-Nagel) and run under electron ionization at 70 eV. The GC was programed with an injection temperature of 250.0 °C (523.15 K) and splitless injection volume of 1.0 µL. For media samples, a 1:20 split injection was used. The GC oven temperature program started at 50 °C (323.15 K) for 1 min, rising to 300.0 °C (573.15 K) at 10 K/min with a final hold at this temperature for 6 min. The GC flow rate with helium carrier gas (HE, HE 5.0UHP, Praxair, Danbury, CT) was 1.2 mL/min. The transfer line temperature was set at 290.0 °C (563.15 K) and ion source temperature at 295.0 °C (568.15 K). A range of 50-600 m/z was scanned with a scan time of 0.25 s.

*Metabolomics data processing*

Metabolites were identified using TraceFinder (v3.3, Thermo Fisher Scientific, Waltham, MA) based on libraries of metabolite retention times and fragmentation patterns (Metaflux, Merced, CA). Identified metabolites were quantified using the selected ion count peak area for specific mass ions, and standard curves generated from reference standards run in parallel. Peak intensities were normalized for extraction efficiency using L-norvaline as an internal standard. The mean and standard deviation for each quantified metabolite was calculated for each cell line and treatment condition. A univariate t-test was used to compare treatment conditions for each metabolite and cell line.

## Results

*Metabolic profiling of a melanoma progression model with activated glutamatergic signaling*

GRM1-positive melanoma patients share a common metabolic dysregulation in glutamatergic signaling. In contrast, the common genotype of melanoma involves mutations of the MAPK pathway, including mutually exclusive mutations of BRAF-, NRAS-, and NF1 [40,41]. In case MAPK driver mutations are lacking, the genotype is classified as NF1/NRAS/BRAF triple negative or triple wild-type subtype and remains without actionable molecular targets. Overcoming the lack of molecular targets of this subtype has been identified as one of the prime goals of the melanoma and pigment cell research community [14]. In order to understand and quantify the metabolic phenotype of GRM1-activated ($GRM1^+$) cells, we profiled an isogenic cellular melanoma model by GCMS (Figure 1).

*Quantifying the intracellular glutamate pool*

The panel includes non-tumorigenic models, C81-61 ($GRM1^{-}$) and C8161si ($GRM1^{KD}$), vs tumorigenic models, C81-61OE ($GRM1^{OE}$) and C8161 ($GRM1^{+}$) (Figure 1A). Several studies have shown that cells expressing GRM1 produce greater amounts of glutamate [10,30-32]. To determine the effect of GRM1 on the intracellular pool size of glutamate, we incubated C81-61, C8161si, C81-61OE, and C8161 melanoma cell lines in glutamate free media and measured intracellular glutamate via gas chromatography mass spectrometry (GCMS). In accordance with previous results, cells with greater GRM1 expression (C81-61OE and C8161) had significantly increased levels of intracellular glutamate than did their counterparts with low GRM1 expression (C81-61 and C8161si) (Figure 1B). Hierarchical clustering of specimens identified two major branches, which resembled tumorigenicity in the melanoma progression model (Figure 1C). C81-61OE ($GRM1^{OE}$) and C8161 ($GRM1^{+}$) showed a cluster characterized by elevated pool sizes of glutamate and tricarboxylic acid (TCA) cycle related metabolites. Among the strongest dysregulated metabolites are α-ketoglutaric acid, aspartic acid, fumaric acid, malic acid, asparagine, trans-4-hydroxyproline, lysine, glutamic acid, and citric acid. There metabolites are significantly elevated in the tumorigenic specimens with p values below $10^{-3}$.

Significantly enriched and associated pathways with this cluster of metabolites include the TCA cycle, glutamate metabolism, aspartate metabolism, as well as arginine and proline metabolism with p values below $10^{-3}$. A principle component analysis converts the set of biochemically possibly correlated metabolite variables into a set of linearly uncorrelated variables called principal components. The principle component analysis plot includes more than 60% data representation in the first two principle components. The multidimensionality reduction plot visualizes the data separation between tumorigenic and non-tumorigenic specimens as well as the metabolic switch and reversal that is induced by the genetic perturbations (Figure 1D). Knockdown of GRM1 in tumorigenic C8161 causes a shift in the metabolic state that is similar to C81-61. In reverse, overexpression of GRM1 in non-tumorigenic C81-61 causes metabolic perturbation into a quadrant of the multidimensional space overlapping with C8161. Taken together, unsupervised clustering and principle component analysis showed metabolic similarity between tumorigenic C8161 (*GRM1*$^{+}$) and C81-61OE (*GRM1*$^{OE}$) and specimens and non-tumorigenic C81-61 (*GRM1*$^{-}$) and C8161si (*GRM1*$^{KD}$). Further, the metabolomic profiling revealed elevated levels of glutamate and TCA cycle-related metabolites in a cancer progression model of malignant melanoma.

*Stable isotope tracing of isogenic cancer progression model shows increased glycolytic flux into glutamate*

Next, we determined the relative contribution of glucose and glutamine to the intracellular glutamate pool by incubating cells with $^{13}C$ labeled tracers and measuring isotopic incorporation into glutamate by GCMS (Figure 2). C8161 (*GRM1*$^{+}$) specimens had elevated pyruvate and lactate production, consistent with the Warburg effect of cancer cells and their proliferative phenotype (Figure 2A). Stable isotope incorporation into the M+3 feature of pyruvate from uniformly labeled ([U-$^{13}C_6$]) glucose indicates increased glycolytic flux in GRM1-activated tumors. While the intracellular pool size increased moderately, the biosynthetic rate increased by more than two-fold with a *p* value below $10^{-3}$. By considering the mass contribution of the carbon source and the tracer-to-tracee ratio, flux into a metabolite can be calculated (Figure 2B). Unexpectedly, while GRM1 expression increased both glycolytic and glutaminolytic flux, the proportional increase in glycolytic glutamate was much higher in C8161 (*GRM1*$^{+}$) cells. Following a 24 hour labeling period with either uniformly labeled $^{13}C_6$-labeled glucose or uniformly labeled $^{13}C_5$-glutamine, cells with higher GRM1 expression had significantly higher levels of $^{13}C$-labeled glutamate from glucose than from cells with low GRM1 expression (Figure 2C). Conversely, cells with high GRM1 expression had a lower percentage of labeled glutamate following incubation with $^{13}C_5$-glutamine (Figure 2D). In summary, stable isotope labeling revealed increased glycolytic flux into glutamate in cells with high GRM1 expression. Interestingly, the total contribution of glutamine to the glutamate pool varied much less with GRM1 expression, indicating that GRM1 expression has a greater impact on glycolytic glutamate production than on glutamate production from glutamine. Taken together, $^{13}C$ labeling in combination with stable isotope tracing identifies glucose and glutamine as distinct carbon sources of oncogenic glutamate, thus providing a rationale for metabolic drug targeting of tumorigenic cancer cells (Figure 2F).

*Pharmacogenomics screen of chemical inhibitors of metabolism*

A pharmacogenomics screen quantified the impact of small molecules on viability and central carbon metabolism in the melanoma progression panel (Table 1, Figure 2F). $IC_{50}$ values, the dose which is lethal to 50% of cells, was determined for the non-tumorigenic C81-61 (*GRM1*$^{-}$) and tumorigenic C8161 (*GRM1*$^{+}$) models (Figure 2G). For each metabolic inhibitor compound, a two-fold dilution series in quadruplicate measurement over a 24-hour incubation period was for assessed for cell viability by a luminescence-based cell viability assay (Figure 2G). Despite the therapeutic window, defined as difference in drug response between cancer and control cell lines, differed greatly for each compound tested, the sensitivity to metabolic inhibitors overall increased with enhanced GRM1 expression.

*Drug sensitivity of GRM1 to glutamate restriction*

Compounds riluzole, BPTES and CB839 target the metabolic proximity and five-carbon supply of glutamate by restricting metabolic exchange of glutamate or conversion from glutamine to glutamate in the GLS reaction [42]. C8161 (*GRM1*$^{+}$) cells were more sensitive to riluzole, a similar response between C81-61 (*GRM1*$^{-}$) and C8161 (*GRM1*$^{+}$) was observed at eight-fold lower concentrations of riluzole, lowering its IC50 value for C8161 (*GRM1*$^{+}$) into the low micromolar range. To test the importance of glutaminolysis in GRM1$^{+}$ cells, we quantified cell viability in the presence of BPTES and CB839, selective inhibitors of GLS. For BPTES a trend of lower cell viability with higher GRM1 expression was noted, yet overall, across any concentrations tested, no significant difference was noted in cell viability between C81-61 (*GRM1*$^{-}$) and C8161 (*GRM1*$^{+}$) cells. Importantly, this result deviated from another more potent GLS inhibitor, CB839, which showed a more than two-fold increase in sensitivity in the tumorigenic cell line and an $IC_{50}$ value in the high nanomolar range (Table 1, Figure 2G).

*Metabolic bottleneck of glycolytic acetyl-CoA for TCA cycle-dependent glutamate production*

Given the increased dependence of tumorigenic GRM1$^{+}$ cells on glycolytic carbon for glutamate production, we next sought to investigate whether inhibiting glycolysis preferentially affects the viability of GRM1$^{+}$ cells. Incubation of cells with 2-deoxy-glucose, a competitive inhibitor of hexokinase and glycolysis, and oxamate, a competitive inhibitor of lactate dehydrogenase (LDH) vital for reduction equivalent regeneration, showed that the tumorigenic cell line is more sensitive to inhibition of glycolysis than the non-tumorigenic parental cell line. Similarly, when incubated with UK5099, a potent inhibitor of the mitochondrial pyruvate transporter (MPC1), the tumorigenic cell line was considerably more sensitive than the non-tumorigenic parental cell line. In addition, the HIF1A modulator, deferoxamine mesylate stronger affected the tumorigenic cell line. These results indicate an increased reliance on the glycolysis in GRM1$^{+}$ cells and that decoupling cytosolic glycolysis from the mitochondrial TCA cycle preferentially affects GRM1$^{+}$ cells.

*Experimental design of preclinical combination trial focused on metabolic signaling and tumor metabolism of malignant melanoma.*

The goal of this preclinical study was to assess alterations in tumor progression following treatment of tumor-bearing mice with metabolic inhibitors vs control cohorts (Figure 3). In this gender-controlled xenograft study, a randomized cohort of male and female animals was injected with either female (C8161) or male (1205LU) tumors. The tumor burden in the presence of control vehicle DMSO, riluzole at 10 mg/kg per os (p.o.), UK5099 3 mg/kg p.o., and CB839 200 mg/kg p.o., as well as combinations of riluzole with UK5099 or CB839 was monitored.

*Significant reduction of tumor burden in combination modality of different metabolic drugs targeting glutamate bioavailability*

Tumor burden in xenograft models was monitored in a gender-controlled study over the time of 28 days, comparing drug single treatment conditions vs combination treatments and control vehicle. Tested drug conditions included single treatments of riluzole, UK5099, CB839, as well as riluzole combined with UK5099 or CB839 (Figure 3). In the C8161 tumor model, the combination modality of riluzole and CB839 performed best, while the combination of riluzole and UK5099 did not improve a reduction of tumor burden (Figure 4A,B). Given these results, in a second large cohort modalities with UK5099 were excluded. The 1205LU xenografts were treated with single treatments of riluzole or CB839, as well as a combination modality of riluzole with CB839. Validating the C8161 tumor xenografts, the 1205LU xenografts responded best to the combination modality with riluzole and CB839 (Figure 4C,D).

*Sex-specific response in tumor xenograft models*

The experimental design included gender-specific xenograft models allowing to assess the impact of cross-sex tumor xenografts (e.g. CB161 tumors into male mice or 1205LU tumors into female mice) in comparison with same-sex tumor xenografts (e.g. CB161 tumors into female mice or 1205LU tumors into male mice). The data showed an overall reduction of the tumor burden under single treatment for cross-sex xenografts. In contrast, the impact of combination treatments in comparison to single treatments was much more pronounced in same-sex treatments. For example, in 1205LU xenografts in female mice, the combination of riluzole and CB839 was comparable to CB839, which already had high efficacy alone. Overall, the comparison of different xenograft systems showed that no single agent treatment was as effective as the combination modality of riluzole with CB839 in reducing tumor progression (Figure 4). Single agent CB839 had the best efficacy, followed by riluzole. In contrast, mice treated with UK5099 as single or combination modality showed the least effectiveness in reducing tumor volumes but rather suffered from weight loss during the study. Also the addition of riluzole to UK5099 did not alter the outcome. In contrast, the combination of riluzole and CB839 significantly improved the reduction of tumor burden over vehicle control and in comparison with single treatments (Figure 4). The preclinical data on tumor burden identified the combination of riluzole and CB839 as the most effective therapeutic strategy tested. Further, results from these in vivo studies uncovered a gender effect in tumor xenograft studies. An enhanced treatment efficacy was observed if donor tumor cells and the gender of the recipient mice match (Figure 4).

*Molecular pathology characterizing cell proliferation, apoptosis, and glutamatergic signaling in excised xenograft tumors*

Immunohistochemistry allowed for assessment of cancer signaling by looking at the overall expression levels of a given marker. The drug-treated tumor tissues showed a significant reduction in protein expression of the oncogenic driver GRM1. In contrast, the vehicle-treated control displays high level of GRM1 protein. Mitogenic signaling pathways associated with MAPK and AKT exhibit reduced activation of key stimulatory factors including phospho mitogen-activated protein kinase 1 (MAPK1/ERK) and phospho AKT. GLS, the protein target of CB839 shows reduced expression in riluzole and the combination modality. In contrast, the single treatment of CB839 displays high levels of GLS comparable to pre-treatment levels, pointing to heterogeneity and potential counter regulation of GLS at the protein level. Histopathology staining for apoptosis and proliferation provided insight into tissue and tumor response under treatment in context of normal and tumor cells. Histopathology staining monitored MKI67 as marker of proliferation of tumor cells and cleaved caspase 3 (CASP3) indicative of apoptosis under drug treatment. Overall, the combination treatment of riluzole with CB839 reduced expression of proliferation markers and pro-survival proteins (Figure 5). The histopathological assessment of the drug trial also included staining of liver sections with hematoxylin and eosin. Histopathology of the liver did not reveal inflamed or necrotic regions of the tissue indicating tolerance of the utilized drug dosages.

*Reduction of plasma amino acid metabolite pool sizes upon combination treatment with GRM1 inhibitor riluzole and glutaminase inhibitor CB839*

Glutamate plasma levels respond to the administered drugs. Circulating glutamate of tumor xenografts is significantly lowered by about 20% on average upon combination treatment with GRM1 inhibitor riluzole and glutaminase inhibitor CB839 with $p$ values below 0.05 (Figure 6). In addition, anaplerotic amino acids and TCA cycle-related organic acids belong to the direct supply chain of biosynthetic tumor glutamate and the tumor microenvironment. Targeted metabolomics revealed that plasma pool sizes of hydroxyl-proline, oxoglutarate, oxalic acid, aspartate, and asparagine showed significant reduction in the combination treatment with $p$ values below $10^{-3}$ (Figure 6). Their profiles indicate a trend of higher responsiveness of plasma metabolite from single treatment with riluzole to single treatment with CB839 and maximum response in the combination modality of both drugs combined.

## Discussion

Tumor models of rewired glutamate metabolism provide a unique focus on noncanonical mitogen-activated protein kinase pathways, G-protein activation independent of NF1/BRAF/NRAS/MAPK genotypes, and metabolic signaling in cancer. In melanoma, hyperactive metabotropic glutamate receptor 1 (GRM1) is an oncogenic driver in the neuroectodermal-derived lineage of melanocytes. Ectopic expression of G protein-coupled receptors have the ability to regulate mitogenic signals [16,43]. A switch to cell cycle progression and biosynthetic metabolism creates overflow metabolism, which in turn produces the extracellular signals required for stimulation of the ectopic transmembrane receptors. GRM1-positive cancer cells release excess glutamate into the extracellular environment, leading to constitutive activation of the receptor. Thus, in melanoma, $GRM1^+$ tumors have acquired a self-stimulatory feedback loop, which is challenging to overcome by pharmacological targeting efforts. GRM1 activation is independent but not mutually exclusive to frequently observed activating BRAF mutations in melanoma [44]. Effective genotype-matched targeting identified the oncogenic event, involves suppression of the signaling cascade, controls biosynthetic production of effector molecules, and break positive enforcement of existing feedback circuits. In detail, the identified combination modality targeting rewired glutamate signaling in melanoma showed reduced expression of the identified upstream G protein-coupled receptor and cancer driver GRM1 itself. Furthermore, it reduced mitogenic mediators phospho-AKT and phospho-ERK [43], suppressed the glutamate-fueling enzyme GLS [44], and restricted overflow metabolism of glutamate into circulation and tumor microenvironment, thereby keeping the effector molecule and feedback trigger, glutamate, in check.

Riluzole and CB839 aim at complementary aspects of glutamate signaling and metabolism. Riluzole has been established as functional inhibitor of the cystine-glutamate antiporter system solute carrier family 7 member 11 (SLC7A11, xCT) preventing release of excess glutamate [45]. In the treatment of amyotrophic lateral sclerosis (ALS), riluzole contains neurotoxic excretion of glutamate [46]. Glutaminase inhibitors are currently evaluated as chemotherapy agents for patients with advanced, metastatic solid and hematopoietic tumors [47,48], where glutamine metabolism has been identified as suitable drug target—even in the absence of hyperactive glutamate biosynthesis. Based on tumor burden and molecular pathology, our data evidence that the combination of glutamate release inhibitor riluzole and GLS inhibitor CB839 is superior to single drug modalities. In the xenografts tested, the combination treatment was tolerated well and even led in few instances to complete elimination of the tumor mass in the absence of systemic toxicities.

The aspect of gender origin of cells utilized in experimental biology has been recognized as important determinant and human cells exhibit wildly different concentrations of many metabolites across the sexes [49,50]. We identified two sex-related aspects in xenograft models contributing to different drug efficacy. The first aspect addresses matching sex between donor tumor specimen and host animal. Potential barriers imposed by immune recognition, cell-to-cell communication, and hormone-dependent signaling may favor elimination of cells with less similar genotypes, leading to reduced engraftment. Tumors injected into a host animal of the opposite gender yielded a significantly reduced tumor burden. A second layer addresses cytotoxicity of a drug, which may be different for each gender depending on differences in xenobiotic metabolism [51]. Further, when looking at an absolute quantity like tumor volume upon drug treatment, such cross-sex related modulation of the engraftment efficacy may lower the reported drug efficacy across a cohort average in the absence of gender stratification. By rigorously implementing gender stratification, personalized treatment regimens are facilitated and reproducibility of disease and developmental models is enabled, consistent with the mandate from the National Institutes of Health to improve scientific rigor and reproducibility in research.

Pharmacogenomics and systems biology suggest that the relationships between the cellular response to drugs, genetic variation of patients and cell metabolism may facilitate managing side effects by personalizing drug prescriptions and nutritional intervention strategies. Genome-wide modeling of drug responses in human flux models have shown that it is possible to predict pathways or closely related metabolic sub-networks responsive to small molecule inhibition [52]. In future, computational prediction will substantially accelerate the drug discovery pipeline. Along the same lines, utilization of *in silico* and cellular models in pharmacogenomics allows high throughput screening and exclusion of less promising drug conditions before entering clinical trials. Still, foreseeing success or failure when translating *in vitro* models to *in vivo* pharmacokinetics is challenging. Our experimental design and rational drug screening pipeline considered metabolic profiling of pathway responses, stable isotope tracing of carbon sources of production metabolites, and plate-based toxicology screens of drug responses. Using stable isotope tracing and GCMS analysis, we determined that cells expressing GRM1 fuel carbon in glutamate from glucose and glutamine. Furthermore, the glycolytic contribution into glutamate is increased compared to normal cells. The increased dependence of glycolytic carbon for glutamate production is further confirmed by the enhanced sensitivity of GRM1 expressing cells to inhibitors that block the anaplerotic flow of carbon into the mitochondrial TCA cycle metabolite pool. UK5099 targeted glycolytically-derived glutamate influx, restrained the bottleneck of mitochondrial pyruvate supply, and significantly reduced the drug response from non-tumorigenic to tumorigenic specimens. Based on outlined criteria, the compound was a promising lead and top *in vitro* performing compound of the drug panel tested. In contrast, treatment with UK5099 did not alleviate the tumor burden in xenografted animals. The fact that mice in the UK5099 treated group exhibited weight loss and behavioral effects indicates a significant bioenergetic impact. However, unexpectedly blockade of the mitochondrial pyruvate carrier eventually benefitted tumor metabolism, selecting for a rewired metabolic state favoring extensive tumor growth and resulting in higher tumor burden instead of reducing it. Notably, further increase of drug dosage was not possible due to apparent side effects.

Our drug targeting strategy considered different aspects of tumor metabolism. The goal of the study was to cut off glutamate supply, while reducing circulating glutamate plasma levels, such that the mitogenic glutamatergic feedback loop can be broken. The combined treatment of riluzole and CB839 leads to enhanced inhibition of $GRM1^+$ melanoma proliferation and reduction of tumor burden *in vivo*. The animal models evidenced significant suppression of tumor progression, in part complete relieve of the tumor, following treatment with the combination of riluzole and CB839. The ability to generate glutamate from glucose confers adaptive advantage to GRM1 activated and dependent cells, particularly those existing in glutamine poor microenvironments. However, *in vivo* experiments showed that limiting glycolytic carbon supply via the mitochondrial pyruvate carrier into the TCA cycle is insufficient. Back-to-back comparison of *in vivo* inhibitor combinations evidenced that inhibition of glutaminolytic carbon has higher efficacy in comparison to glycolytic carbon sources.

By following the flow of carbon from glucose and glutamine in GRM1 expressing melanoma, we have identified fundamental metabolic pathways that support autocrine glutamatergic signaling. Rational targeting of these pathways therefore provides novel combinatorial treatment focused on glutaminases and glutamate release inhibitors in the treatment of GRM1 expressing malignancies.

## Conclusion

Stable isotope tracing and GCMS analysis revealed that expression of the metabotropic glutamate receptor GRM1 leads to an increase in the production of glutamate in melanoma cells. Though it has been thought that increased glutamine uptake and glutaminolysis is responsible for increased glutamate production, our results indicate an important role for glycolysis and glucose derived glutamate in cells expressing GRM1. In fact, while GRM1 expression increases the overall amount of glutamate produced the proportion of glucose versus glutamine derived glutamate is increased. This data was supported by an increased sensitivity to inhibitors of glycolytic metabolism in cells over expressing GRM1. Together, these results suggest that targeting increased glutamate production from glutaminolytic and glycolytic sources may be an effective treatment for GRM1 expressing melanoma.

## Declarations

### Ethics approval and consent to participate

Not applicable.

### Competing interests

There is no competing financial interest.

### Acknowledgements

F.V.F. is grateful for the support by grants CA154887 from the National Institutes of Health, National Cancer Institute, GM115293 NIH Bridges to the Doctorate, NSF GRFP Graduate Research Fellowship Program, CRN-17-427258 by the University of California, Office of the President, Cancer Research Coordinating Committee, and the Science Alliance on Precision Medicine and Cancer Prevention by the German Federal Foreign Office, implemented by the Goethe-Institute, Washington, DC, USA, and supported by the Federation of German Industries (BDI), Berlin, Germany. S.C. thanks for grant from New Jersey Health Foundation and Veterans Administration Research award 101BX003742. R.S. is thankful for a Bristol-Myers Squibb Ph.D. Fellowship. S.C. and R.S. are grateful for the support of the NIH NIEHS T32 Training Grant ES007148 in Environmental Toxicology and the New York Society of Cosmetic Chemists. We would like to thank Muhammad Karabala for assistance with pharmacogenomics screening and metabolite extraction, and Darling Rojas and Kevinn Eddy for support with animal and biochemical work.

## Figures

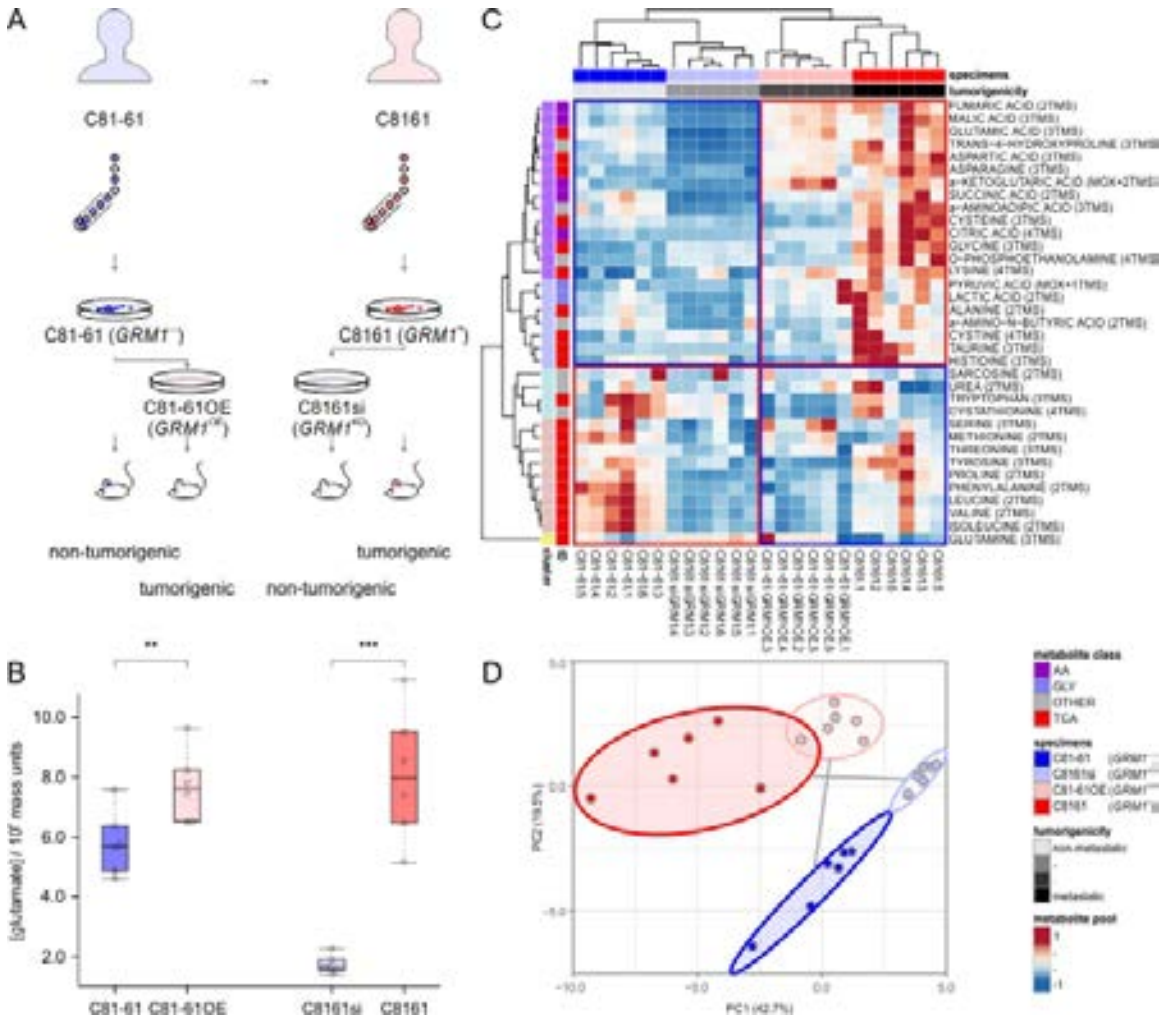

**Figure 1:** Metabolomic profiling of isogenic cancer progression model

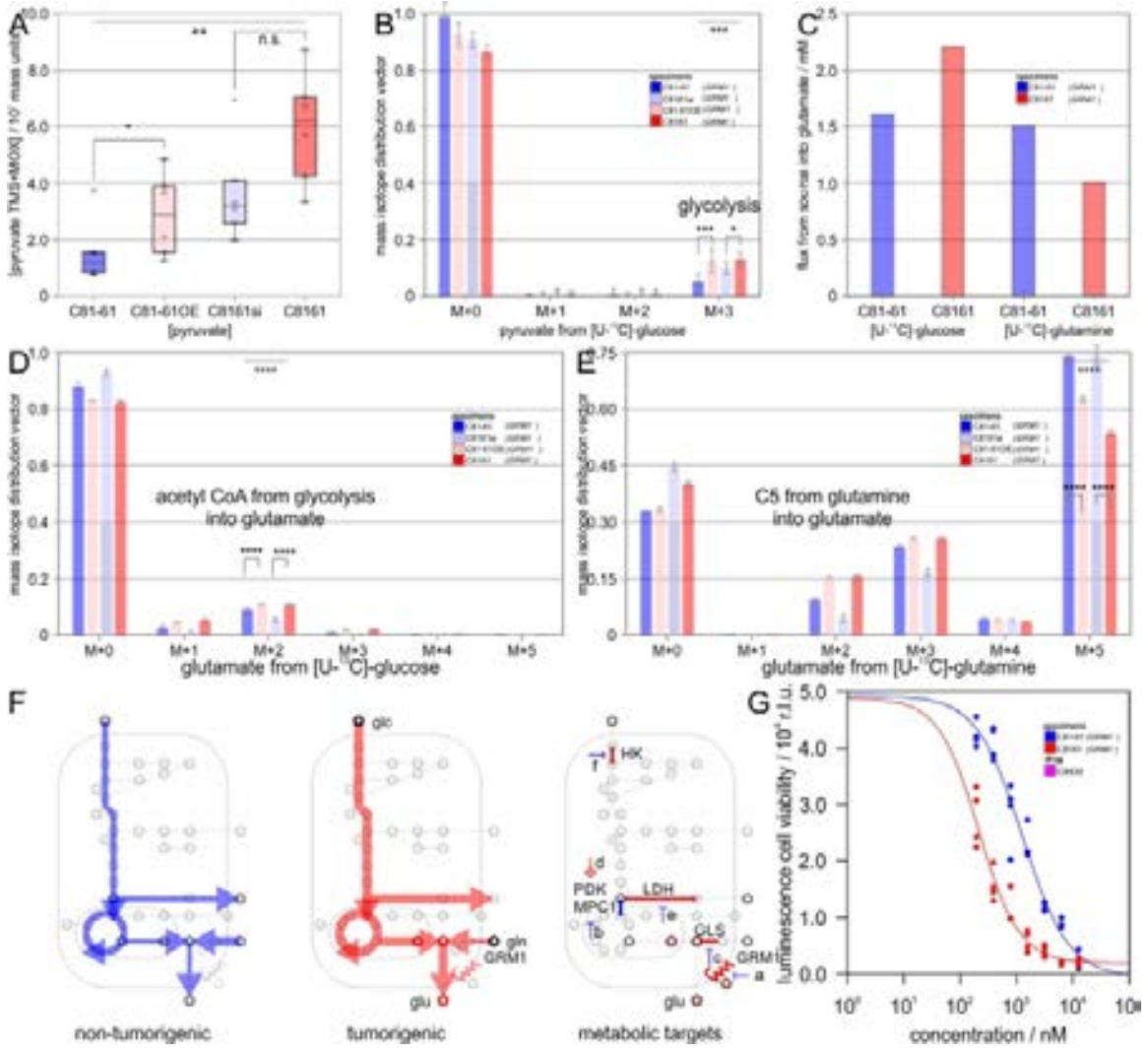

**Figure 2:** Stable isotope tracing identifies dual carbon source and directs metabolic targeting of tumorigenic cancer cells

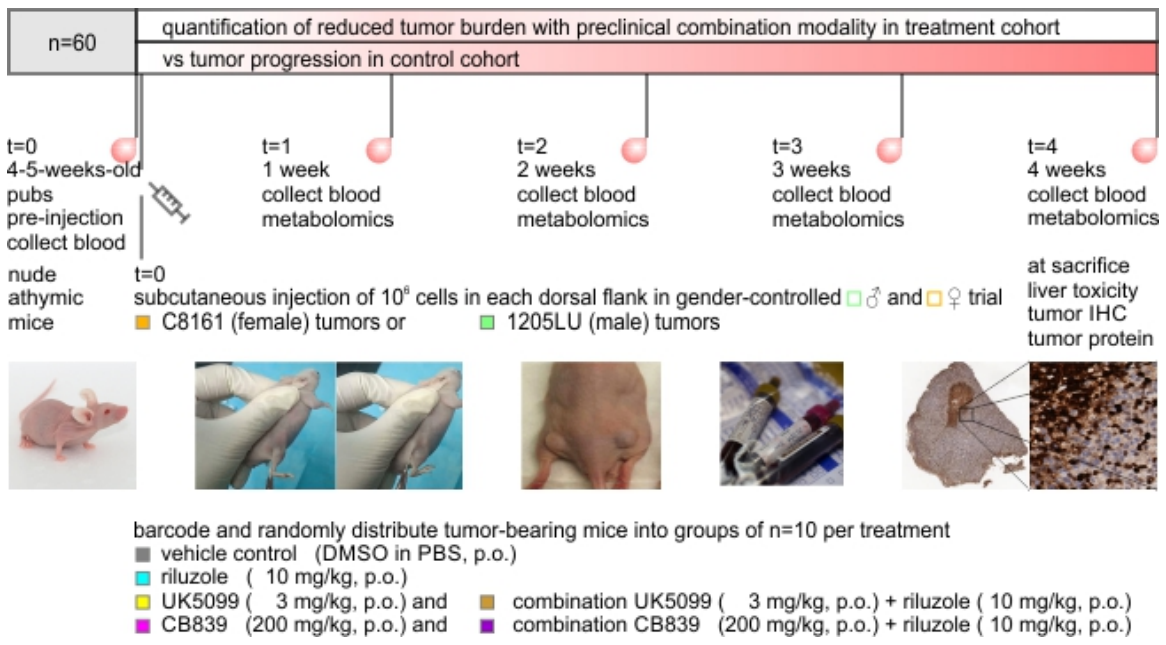

**Figure 3:** Preclinical combination trial targets metabolic signaling and tumor metabolism of malignant melanoma

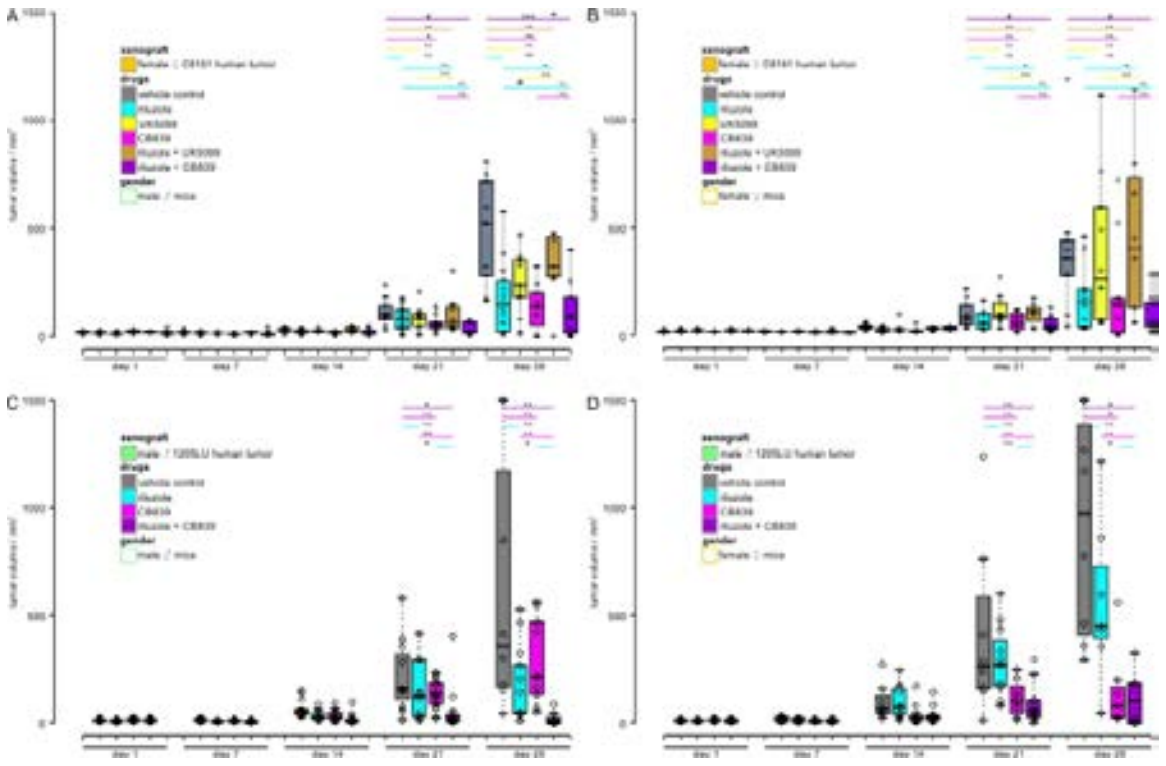

**Figure 4:** The combination modality of riluzole and CB839 significantly reduces tumor burden vs vehicle control or single treatment

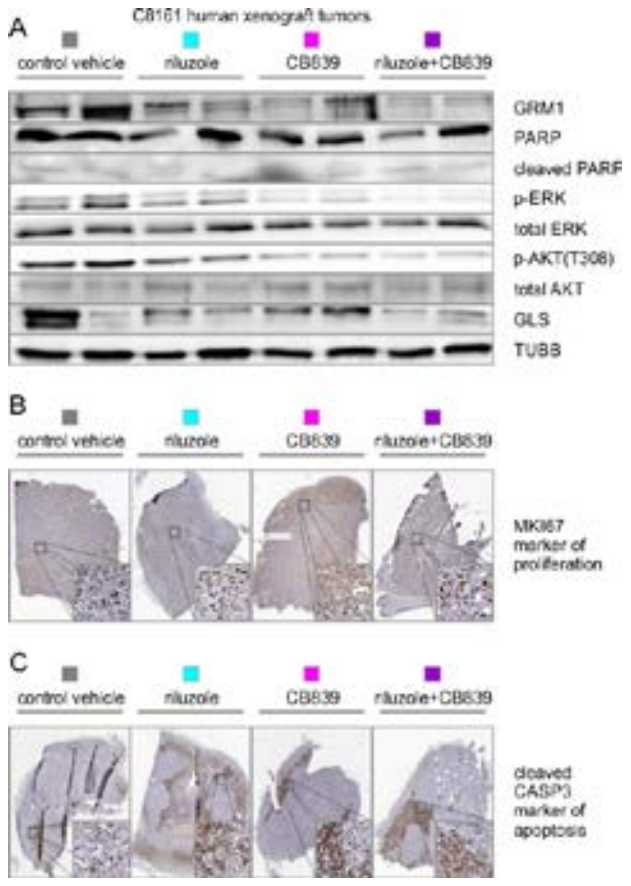

**Figure 5:** Molecular characterization of tumor xenografts under drug treatment

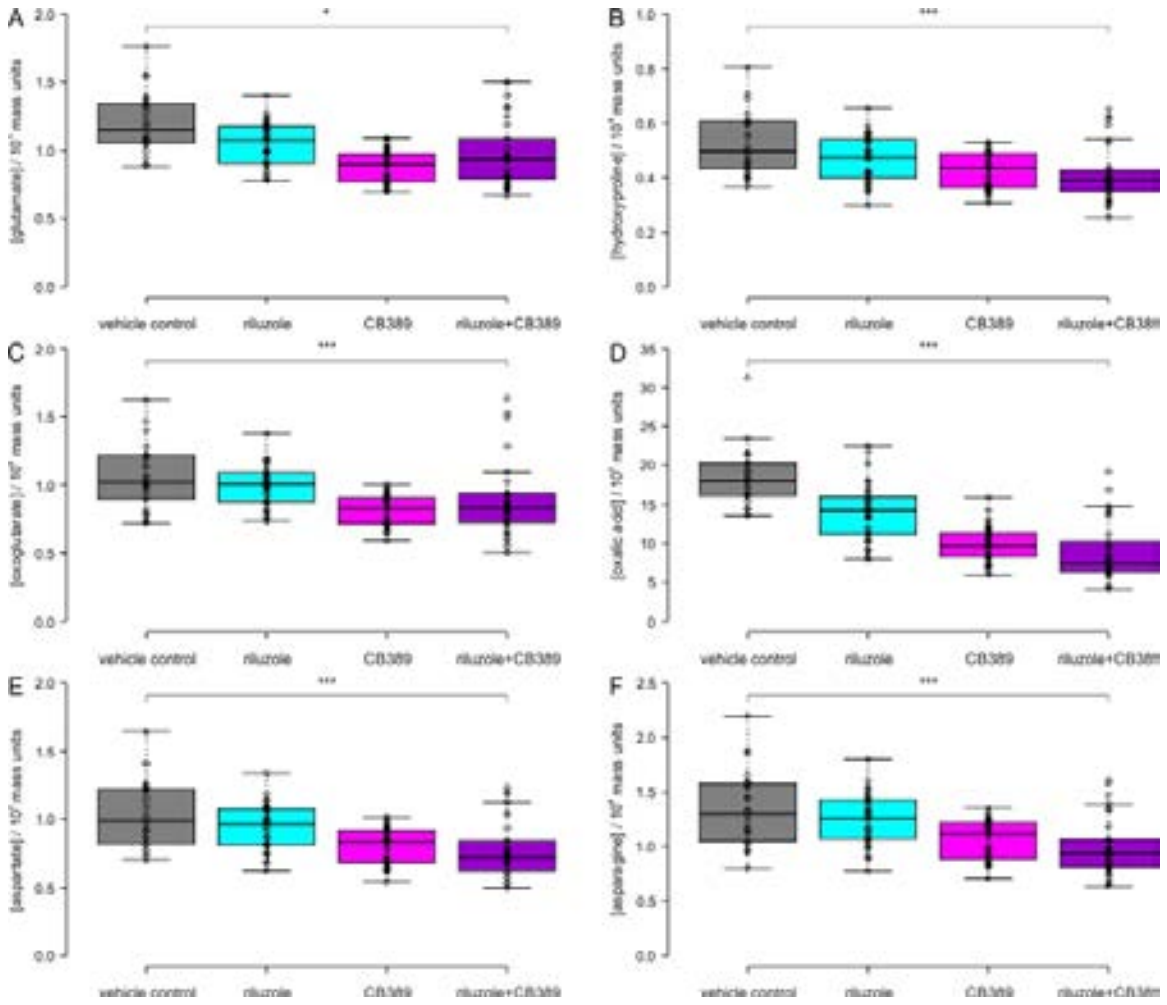

**Figure 6:** Reduction of circulating metabolite levels in combination modality of riluzole and CB839 targeting GRM1 activated melanoma

**Table 1:** Pharmacogenomics screen targeting rewired metabolism in melanoma.

| compound | metabolic effect | IC50 non-tumorigenic C81-61 (*GRM1*⁻) | IC50 tumorigenic C8161 (*GRM1*⁺) |
|---|---|---|---|
| *a = riluzole* | inhibits GRM1 | 152.03 µM | 19.50 µM |
| *b = BPTES* | inhibits glutaminase | 9.55 µM | 5.98 µM |
| *c = CB839* | inhibits glutaminase | 505.0 nM | 222.0 nM |
| *d = UK5099* | inhibits mito pyruvate carrier | 420.0 nM | 4.7 nM |
| *e = dicloroacetate* | inhibits PDK, activates PDH | 113.068 mM | 114.665 mM |
| *f = oxamate* | inhibits LDH | 40.027 mM | 9.133 mM |
| *g = 2-deoxyglucose* | inhibits HK | 5.518 mM | 3.796 mM |
| *h = deferoxamine mesylate* | stabilizes HIF1A | 161.19 µM | 142.93 µM |